\documentclass[showpacs,twocolumn,floats,superscriptaddress]{revtex4}
\usepackage{graphicx}
\usepackage{amsmath}
\begin{document}
\title{Quantum Andreev map: A paradigm of quantum chaos in
superconductivity}
\author{Ph. Jacquod}
\affiliation{Instituut-Lorentz, Universiteit Leiden, P.O. Box 9506, 2300 RA
Leiden, The Netherlands}
\author{H. Schomerus}
\affiliation{Max Planck Institute for the Physics of Complex Systems,
N\"{o}thnitzer Str.\ 38, 01187 Dresden, Germany}
\author{C.W.J. Beenakker$^1$}
\date{March 2003}
\begin{abstract}
We introduce quantum maps with particle-hole conversion
(Andreev reflection) and particle-hole symmetry, which exhibit the same
excitation gap as quantum dots in the proximity to a superconductor.
Computationally, the Andreev maps are much more efficient than billiard
models of quantum dots. This makes it possible to test analytical
predictions of random-matrix theory and semiclassical chaos
that were previously out of reach of computer simulations.
We have observed the universal distribution of the excitation gap 
for large Lyapunov exponent and the logarithmic reduction of the gap when the
Ehrenfest time becomes comparable to the quasiparticle dwell time.
\end{abstract}
\maketitle

One-dimensional (1D) quantum mechanical models with a chaotic classical
limit were first studied by Casati, Chirikov, Ford, and Izrailev in
1979 \cite{Cas79}. These models
have since
developed into one of two paradigms
of quantum chaos \cite{Izr90,Haa92}. The other paradigm is the 2D
billiard of irregular shape \cite{Boh84}. Two is the lowest number of dimensions for
non-integrable (chaotic) dynamics in autonomous systems,
since a single constant of motion is sufficient for integrability
in 1D. The 1D models get around this constraint through a periodically
time-dependent external force (``kick''), which eliminates the energy
as constant of motion --- but still conserves the quasi-energy
(analogously to quasi-momentum conservation in a periodic lattice).
The two paradigms share a common set of phenomena in the fields
of quantum chaos and localization \cite{Fis82,Alt96,Bor96,Fra97}.

The combination of chaos and superconductivity produces an entirely
new phenomenology, notably the appearance of an excitation gap as a
signature of quantum chaos \cite{Mel96}. The paradigm common to most
of the literature is the 2D billiard connected to a superconductor
\cite{Gut02},
introduced under the name ``Andreev billiard'' in Ref.\ \cite{Kos95}. The
name refers to the Andreev reflection which occurs at the interface with
the superconductor, where an electron at energy $\varepsilon$ above the
Fermi level is converted into a hole at energy $\varepsilon$ below it.

From the point of view of computational efficiency, compact quantum maps such
as the kicked rotator \cite{Cas79} (a particle confined
to a circle and driven periodically in time with a strength that
depends on its position) are much more powerful than 2D models such as
billiards. Indeed, there exists a highly efficient
diagonalization technique that works only for maps \cite{Ket99}. The
lack of a 1D map for quantum chaos with superconductivity has
hindered the numerical test of a variety of analytical predictions
\cite{Vav01,Ost01,Lod98,Lam01,Tar01,Ada02,Sil02,Vav02}. Most notably,
numerical efforts have not been able to distinguish the conflicting
predictions \cite{Mel96} of random-matrix theory (RMT) and
the semiclassical Bohr-Sommerfeld (BS) quantization:
RMT predicts an
excitation gap at the Thouless energy while BS gives an exponentially
vanishing density of states without a true gap.  Recent analytical work
\cite{Tar01,Ada02,Sil02,Vav02,Lod98} has predicted that diffraction creates a
gap in the BS density of states at the inverse Ehrenfest time. This
has never been seen in computer simulations, because the Ehrenfest time
scales logarithmically with the system size and is usually far too small to
play a role.
For these reasons there is a real need for something like a ``quantum Andreev
map''. Does it exist? And if it does, can it be simulated more
efficiently than the Andreev billiard? These are the issues addressed
in this paper.

We show how to construct quantum Andreev maps out of any conventional
quantum map (not necessarily chaotic), in much the same way as any
normal billiard can be turned into an Andreev billiard by coupling it to
a superconductor. The construction is guided by the classical electron
and hole dynamics on the Poincar\'{e} surface of section of an Andreev
billiard. The Andreev kicked rotator is a particular example of such
an Andreev map. We certify that it possesses the phenomenology 
of the Andreev billiards and search for predictions of RMT and semiclassics.
We leave for future investigations the application of the Andreev map to
other kicked models (possibly with a different phenomenology), such
as the kicked top \cite{Haa92} and the Fermi-Ulam model \cite{Jain93}.

A quantum map is represented by the Floquet operator $F$, which
gives the stroboscopic time evolution $u(p\tau_0)=F^{p}u(0)$ of
an initial wave function $u(0)$. (We set the stroboscopic period
$\tau_0=1$ in most equations.)  The unitary $M\times M$ matrix $F$
has eigenvalues $\exp(-i\varepsilon_{m})$, with the 
quasi-energies $\varepsilon_{m}\in
(-\pi,\pi)$ (measured in units of $\hbar/\tau_{0}$).
This describes particle excitations in a normal metal.  We also need
hole excitations. A particle excitation with energy $\varepsilon_{m}$
(measured relatively to the Fermi level) is identical to a hole excitation
with energy  $-\varepsilon_{m}$.  This means that hole excitations
in a normal metal have Floquet operator $F^{\ast}$ and wave function
$v(p)=(F^{\ast})^{p}v(0)$.  

Particles and holes are coupled by connecting the normal metal via a lead
to a superconducting reservoir. On a Poincar\'{e} surface of section,
the lead is represented by a spatially localized region in which Andreev
reflection converts electrons into holes and vice versa, with phase
shift $-i$. (A weak energy dependence of this phase shift is ignored for
simplicity, but can be accounted for straightforwardly.) Analogously,
for the quantum Andreev map we assume that Andreev reflections occur
whenever an excitation ends up in a certain subspace of Hilbert space.
This subspace $n_{1},n_{2},\ldots n_{N}$ consists of $N$ out of $M$ states
in a chosen basis and corresponds to a lead with $N$ propagating channels.
The $N\times M$ matrix $P$ projects onto the lead. Its elements are
$P_{nm}=1$ if $m=n \in \{n_{1},n_{2},\ldots n_{N}\}$ and $P_{nm}=0$ otherwise.
The dwell time of a quasiparticle excitation
in the normal metal is $\tau_{\rm dwell}=M/N$, 
equal to the mean time between Andreev reflections. The fact that 
Andreev reflections only
occur at multiples of the stroboscopic time $\tau_0$ is technically convenient,
and should be physically irrelevant for $\tau_{\rm dwell} \gg \tau_0$.

Putting all this together we construct the quantum Andreev map from the
matrix product
\begin{equation}
{\cal F}={\cal P}
\left(\begin{array}{cc}
F&0\\
0&F^{\ast}
\end{array}\right),\;\;
{\cal P}=\left(\begin{array}{cc}
1-P^{\rm T}P&-iP^{\rm T}P\\
-iP^{\rm T}P&1-P^{\rm T}P
\end{array}\right).\label{calFdef}
\end{equation}
(The superscript ``T'' indicates the transpose of a matrix.) The
particle-hole wave function $\Psi=(u,v)$ evolves in time as $\Psi(p)={\cal
F}^{p}\Psi(0)$. The Floquet operator can be symmetrized (without changing
its eigenvalues) by the unitary transformation ${\cal F}\rightarrow
{\cal P}^{-1/2}{\cal F}{\cal P}^{1/2}$, with
\begin{equation}
{\cal P}^{1/2}=\left(\begin{array}{cc}
1-(1-{\textstyle\frac{1}{2}}\sqrt{2})P^{\rm T}P&-i{\textstyle\frac{1}{2}}\sqrt{2}P^{\rm T}P\\
-i{\textstyle\frac{1}{2}}\sqrt{2}P^{\rm T}P&1-(1-{\textstyle\frac{1}{2}}\sqrt{2})P^{\rm T}P
\end{array}\right).\label{calFsymdef}
\end{equation} 

In order to establish the correspondence of the 1D quantum Andreev maps to
2D Andreev billiards, we examine the spectral properties of the map.
The Floquet operator ${\cal F}$ possesses a particle-hole symmetry
which entails that its $2M$ eigenvalues $\exp(-i\varepsilon_{m})$ come in
inverse pairs. This symmetry is the analogue of the particle-hole symmetry
in Andreev billiards, in which excitation energies $\pm\varepsilon$ occur
symmetrically around the Fermi level.  To avoid double-counting of levels,
we restrict the quasi-energy to the interval $(0,\pi)$.  The excitation
spectrum of particles and holes consists of the $M$ quasi-energies in
this interval, and the mean level spacing $\pi/M$ is twice as small as
the level spacing $\delta=2\pi/M$ for particles and holes separately.
The energy scale for the proximity-induced excitation gap is the
Thouless energy $E_{\rm
T}=N\delta/4\pi=N/2M=1/(2\tau_{\rm dwell})$.

The quantization condition ${\rm det}({\cal F}-e^{-i\varepsilon})=0 $
can be written equivalently as
\begin{equation}
{\rm det}[1+S(\varepsilon)S^{\ast}(-\varepsilon)]=0,\label{detS}
\end{equation}
in terms of the $N\times N$ scattering matrix \cite{Oss02}
\begin{equation}
S(\varepsilon)=P[e^{-i\varepsilon}-F(1-P^{\rm T}P)]^{-1}
FP^{\rm T}.\label{Sdef}
\end{equation}
Eq.\ (\ref{detS}) for the Andreev map has the same form as for the
Andreev billiard \cite{Fra96}, but there $S$ is given in terms of a
Hamiltonian $H_{0}$ instead of a Floquet operator $F$. The two
approaches become entirely equivalent in the context of RMT, when $H_{0}$
is chosen from one of the Gaussian ensembles and $F$ is chosen from
one of the circular ensembles \cite{proof}. They are also equivalent
in the semiclassical limit, when the billiard can be represented by a
Poincar\'{e} map which can be quantized approximately \cite{Bog92}.

In the mean-field limit $M\gg N\gg 1$, RMT predicts a
hard gap in the excitation spectrum of size $E_{\rm RMT}=\gamma
E_{\rm T}$ (with $\gamma=2^{-3/2}(\sqrt{5}-1)^{5/2} = 0.60$),
and above the gap a square-root singularity in the density of
states $\rho(\varepsilon)=\pi^{-1}\Delta^{-3/2}(\varepsilon-E_{\rm
RMT})^{1/2}$ (with $\Delta=0.068\,N^{1/3}\delta$) \cite{Mel96}. Sample-to-sample
fluctuations of the lowest excitation energy $\varepsilon_{0}$ around
the mean-field gap have been calculated in Refs.\ \cite{Vav01,Ost01}. A
universal probability distribution was predicted for the rescaled energy
$x=(\varepsilon_{0}-E_{\rm RMT})/\Delta$. While the mean-field prediction
of RMT has been tested numerically in an Andreev billiard \cite{Mel96},
the numerical error bars
are too large to 
extract the predicted universal
gap fluctuations.

To demonstrate the efficiency of the quantum Andreev maps,
we specialize to the quantum kicked rotator. The
Floquet operator is \cite{Izr90}
\begin{eqnarray}
F&=&\exp\left(i\frac{\hbar\tau_{0}}{4I_{0}}\frac{\partial^{2}}{\partial\theta^{2}} \right)\exp\left(-i\frac{KI_{0}}{\hbar\tau_{0}}\cos\theta\right)\nonumber\\
&&\mbox{}\times\exp\left(i\frac{\hbar\tau_{0}}{4I_{0}}\frac{\partial^{2}}{\partial\theta^{2}} \right), \label{kickedF}
\end{eqnarray}
with $I_{0}$ the moment of inertia of the particle and $K$ the
(dimensionless) kicking strength. The particle moves freely along the
circle for half a period, is then kicked with a strength $K\cos\theta$,
and proceeds freely for another half period. The transition from
classical to quantum behavior is governed by the effective Planck
constant $\hbar_{\rm eff}\equiv\hbar\tau_{0}/I_{0}$. 
Since we would like to compare the kicked rotator to a chaotic billiard,
without localization, we follow the usual procedure of quantizing phase
space on the torus $\theta,p\in(0,2\pi)$, 
rather than on a cylinder, with
$p=-i\hbar_{\rm eff}\partial/\partial\theta$, the dimensionless angular
momentum \cite{Izr90}. For $\hbar_{\rm eff}=2\pi/M$, with integer $M$, the
Floquet operator is an $M\times M$ unitary symmetric matrix. In 
angular momentum
representation it has elements
\begin{subequations}
\label{eq:rotator}
\begin{eqnarray}
&&F_{kk'}=e^{-(i\pi/2M)(k^{2}+k'^{2})}(UQU^{\dagger})_{kk'},\label{Fkkdef}\\
&&U_{kk'}=M^{-1/2}e^{(2\pi i/M)kk'},\label{Udef}\\
&&Q_{kk'}=\delta_{kk'}e^{-(iMK/2\pi)\cos(2\pi k/M)}.\label{Qdef}
\end{eqnarray}
\end{subequations}
Upon increasing $K$ the classical dynamics varies from fully
integrable ($K=0$) to fully chaotic [$K\agt 7$, with Lyapunov exponent
$\lambda\approx\ln (K/2)$]. For $K<7$ stable and unstable motion coexist
(a so-called mixed phase space). 

To introduce the Andreev reflection we use a projection operator which
is diagonal in $p$-representation, 
\begin{equation}
(P^{\rm T}P)_{kk'}=\delta_{kk'}\times\left\{\begin{array}{l}
1\;\;{\rm if}\;\;L\leq k\leq L+N-1,\\
0\;\;{\rm otherwise.}
\end{array}\right. \label{PPTdef}
\end{equation}
(We checked that
similar results are obtained when $P$ is diagonal in $\theta$-representation.)
The position $L$ of the lead to the superconductor is arbitrary. The
Floquet operator ${\cal F}$ of the ``Andreev kicked rotator'' is then
obtained by inserting Eqs.\ (\ref{eq:rotator}) and (\ref{PPTdef}) into Eq.\ (\ref{calFdef}).
We apply the symmetrization (\ref{calFsymdef}), so ${\cal F}$ is a unitary
symmetric matrix. The real symmetric matrix $\frac{1}{2}({\cal F}+{\cal
F}^{\dagger})$ can be diagonalized efficiently with ${\cal O}(M^2 \ln M)$ 
operations (and not ${\cal O}(M^3)$ as with standard methods) by means of 
the Lanczos
technique, if the multiplication with the matrix $U$ is carried out
with the help of the Fast-Fourier-Transform algorithm \cite{Ket99}. The
eigenvalues $\cos\varepsilon_{m}$ uniquely determine the quasi-energy
$\varepsilon_{m}\in(0,\pi)$. 

\begin{figure}
\includegraphics[width=8cm]{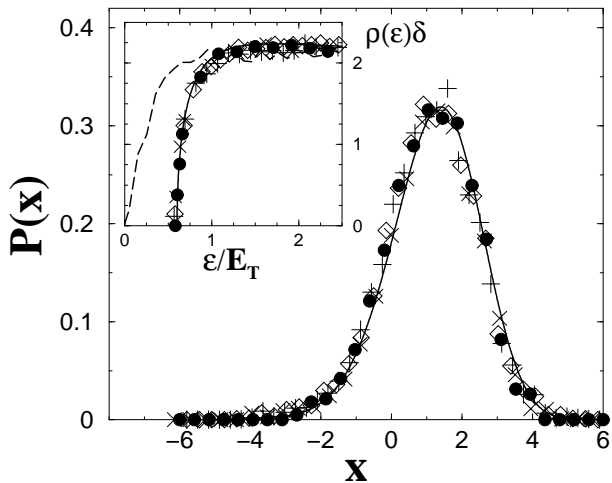}
\\[-3mm]
\caption{Main plot: Gap distribution for the Andreev kicked rotator 
with $M=8192$, $K=45$,
and $M/N=\tau_{\rm dwell}=10$ (diamonds), 
20 (circles), 40 (+), and 50 ($\times$).
The solid line gives the RMT prediction \cite{Vav01}.
Inset: Density of states for the same system.
The solid line is the RMT prediction \cite{Mel96}. The dashed line
is a numerical result in the mixed regime ($M=8192$, $K=1.2$,
$M/N=10$).}  
\label{figure1}
\end{figure}

In the inset to
Fig.\ \ref{figure1} we show the density of states $\rho(\varepsilon)$
for system size $M=8192$, kicking strength $K=45$ (strongly chaotic
dynamics), and several widths $N$ of the lead to the superconductor. The
density of states has been averaged over 250 different positions of the
lead. The data points fall on top of the RMT prediction \cite{Mel96}
without any adjustable parameter. Reducing the kicking strength down to
$K=1.2$, one enters the regime of mixed classical dynamics. 
We see that the gap disappears, as predicted in Ref.\ \cite{Sch99}.

\begin{figure}
\includegraphics[width=8cm]{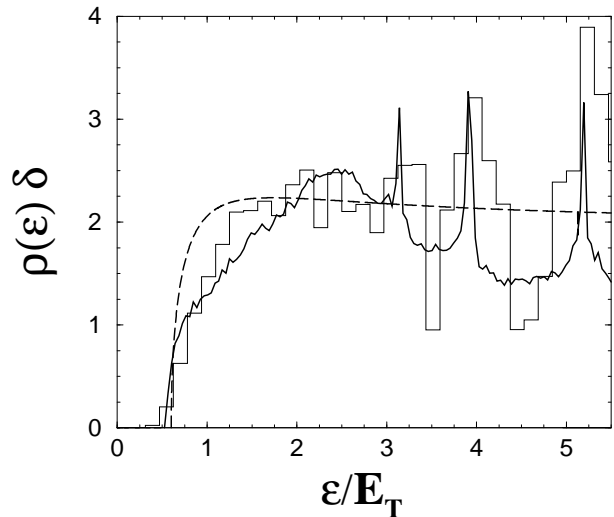}
\caption{Density of states for the Andreev kicked rotator
with $M=131072$, $\tau_{\rm dwell}=5$, and $K=14$ (solid line), compared with
the Bohr-Sommerfeld calculation (histogram), and the RMT prediction (dashed 
line).}  
\label{figure2}
\end{figure}

To test RMT beyond the mean-field limit, we study the statistical
fluctuations of the gap. The main panel of 
Fig.\ \ref{figure1} shows the probability
distribution of the smallest eigenvalue $\varepsilon_{0}$
in the chaotic regime. To improve statistics we
sampled 6000 different positions of the lead. We rescaled the energy
$x=(\varepsilon_{0}-E_{\rm RMT})/\Delta$, as prescribed by Ref.\
\cite{Vav01}. Good agreement is observed with the universal scaling
distribution \cite{Tra94}, again without any adjustable parameters.

It is predicted theoretically that deviations from RMT should appear
if the Ehrenfest time $\tau_{\rm E} = \lambda^{-1} \ln M$ is no longer small
compared to $\tau_{\rm dwell}$. For $\tau_{\rm E} \gtrsim \tau_{\rm dwell}$
the semiclassical Bohr-Sommerfeld approximation \cite{Mel96,Sch99} should
be valid, with a diffraction induced gap of order 
$\hbar/\tau_{\rm E}$ \cite{Lod98}. To search for these deviations
from RMT we consider rotators 
with smaller kicking strengths (but still in the fully chaotic regime), 
thus smaller Lyapunov exponent, and much larger $M$.

In Fig.\ \ref{figure2} we show the density of states for $M=131072$
and $K=14$. Strong deviations from the RMT prediction
are clearly visible. 
Also plotted is the result of a BS calculation \cite{Mel96},
in which we slightly smoothed the singular delta functions.
This approximation agrees better with the exact result. Most remarkably, it reproduces
the three distinct peaks in the density of states, which now
can be identified with trajectories of certain lengths.
All trajectories with lengths that are odd multiples of 
$\tau_{\rm dwell}=5$ contribute to the peak at $\varepsilon/E_{\rm T}=\pi$,
odd multiples of $4$ contribute at $\varepsilon/E_{\rm T}=5 \pi/4$,
and odd multiples of $3$ contribute at $\varepsilon/E_{\rm T}=5 \pi/3$.

\begin{figure}
\includegraphics[width=8cm]{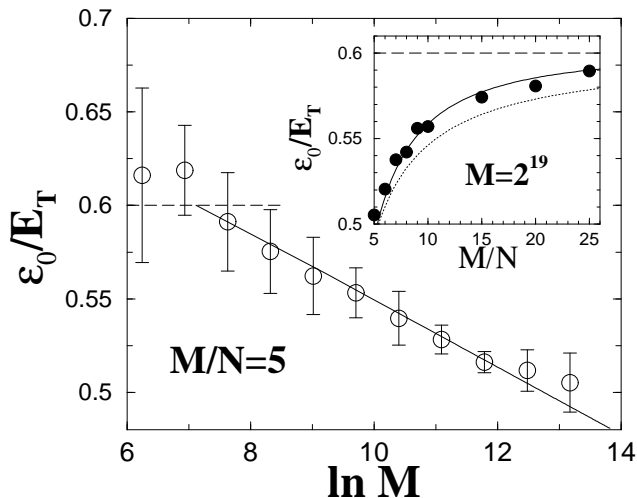}
\caption{Main plot: Dependence of the mean gap on the system size $M$, for 
$\tau_{\rm dwell}=M/N=5$ and $K=14.$ Averages have been calculated
with 400 (for $M=512$) to 40 (for $M > 5 \cdot 10^5$) different positions
of the contacts to the superconductor. The error bars represent the
root-mean-square of $\varepsilon_0$. The dashed line is the 
RMT prediction and the solid line is a linear fit to the data points.
Inset: 
Dependence of the mean gap on $\tau_{\rm dwell}=M/N$ for $K=14$ and
$M=524288$. The dashed line is the RMT prediction and the solid and
dotted curves are given by Eqs.\ (\ref{gapfction}) and (\ref{gapfction2}),
respectively, with coefficients extracted from the linear fit
in the main plot.\\[-8mm]}  \label{figure3}
\end{figure}

A systematic reduction of the excitation gap 
is observed upon increasing the ratio $\tau_{\rm E}/\tau_{\rm dwell}$,
as shown in Fig.\ \ref{figure3}. The main panel is a semi-logarithmic
plot of $\varepsilon_0/E_T$ as a function of $M \in [2^9,2^{19}]$,
for $M/N=\tau_{\rm dwell}=5$ and
$K=14$. Existing theories \cite{Vav02,Sil02} predict a linear initial decrease 
of $\varepsilon_{0}$ with $\ln M$ 
at fixed $\tau_{\rm dwell}=M/N$. We fit the data to the prediction of 
Vavilov and Larkin \cite{Vav02}, 
\begin{equation}\label{gapfction}
\frac{\varepsilon_0}{E_{\rm RMT}}=1-
\frac{\alpha}{2 \lambda \tau_{\rm dwell}}
\left(\ln M-2\ln \frac{M}{N} - \alpha' \right).
\end{equation}
We find $\alpha=0.59$ and $\alpha'=3.95$. Once 
$\alpha$ and $\alpha'$ are extracted, no free parameter is left,
and the resulting curve, shown with a solid line
in the inset to Fig.\ \ref{figure3}, correctly reproduces the
parametric dependence on $\tau_{\rm dwell}=M/N$ at fixed $M$. 
As a further check, we tried the slightly different expression
\begin{equation}\label{gapfction2}
\frac{\varepsilon_0}{E_{\rm RMT}}=1-
\frac{\alpha}{2 \lambda \tau_{\rm dwell}}
\left(\ln M- \alpha'' \right),
\end{equation}
with $\alpha''=\alpha'+2 \ln 5$. The resulting curve (dotted line in the 
inset to Fig.\ \ref{figure3}) shows significant deviations from 
the numerical data.
We conclude that Eq.\ (\ref{gapfction}) gives the correct parametric 
dependence of the Andreev gap. 

A discrepancy remains in the value of the numerical coefficients.
While the coefficient $\alpha'$ is model dependent, the prefactor $\alpha$
is expected to be universal. Our numerics gives $\alpha=0.59 \pm 0.08$,
in between the two competing predictions $\alpha=0.23$ of Ref.\ \cite{Vav02}
and $\alpha=2$ of Ref. \cite{Sil02}.

In conclusion, we have constructed a quantum map
that accounts for the presence of superconductivity. The ``Andreev kicked
rotator'' introduced above, has been shown to be
equivalent to the Andreev billiards studied so far. Owing
to the fact that it is one-dimensional rather than two-dimensional, 
it is much more efficient computationally, which permits
to observe two theoretical predictions that are currently out of reach
of billiard simulations: The universal gap fluctuations for large
Lyapunov exponent
and the logarithmic reduction of the gap for small Lyapunov exponent.
We foresee that the Andreev kicked rotator on a cylinder (instead of on
a torus) can be an equally effective tool to study the interplay of
superconductivity and localization.

We have benefitted from discussions with \.I. Adagideli and
J. Tworzyd{\l}o.
This work was supported by the Dutch Science Foundation NWO/FOM.

\end{document}